\documentclass{article}

\usepackage{microtype}
\usepackage{graphicx}
\usepackage{subfigure}
\usepackage{booktabs} % for professional tables

\usepackage{hyperref}

\usepackage[accepted]{icml2023}

\usepackage{amsmath}
\usepackage{amssymb}
\usepackage{mathtools}
\usepackage{amsthm}

\usepackage[capitalize,noabbrev]{cleveref}

\theoremstyle{plain}
\newtheorem{theorem}{Theorem}[section]

\newtheorem{lemma}[theorem]{Lemma}

\theoremstyle{definition}

\newtheorem{assumption}[theorem]{Assumption}
\theoremstyle{remark}
\newtheorem{remark}[theorem]{Remark}

\usepackage[textsize=tiny]{todonotes}

\icmltitlerunning{Sub-linear Regret in Adaptive Model Predictive Control}

\usepackage{comment}
\def\NN{{\mathbb N}}
\def\EE{{\mathbb E}}

\def\RR{{\mathbb R}}

\newcommand{\ep}{\hfill $\Box$}

\newcommand{\set}[1]{\mathcal{#1}}
\newcommand{\norm}[1]{\lVert#1\rVert}

\begin{document}

\twocolumn[
\icmltitle{Sub-linear Regret in Adaptive Model Predictive Control}

% submissions: thehttps://www.overleaf.com/project/6376013355a8a92cb159ec13 style file will automatically remove it for you


\icmlsetsymbol{equal}{*}

\begin{icmlauthorlist}
\icmlauthor{Damianos Tranos}{KTH}
\icmlauthor{Alexandre Proutiere}{KTH}
\end{icmlauthorlist}

\icmlaffiliation{KTH}{Division of Decision and Control Systems, KTH Royal Institute of Technology, Stockholm, Sweden}

\icmlcorrespondingauthor{Damianos Tranos}{tranos@kth.se}

\icmlkeywords{Regret Analysis, Model Predictive Control, Adaptive Control}

\vskip 0.3in
]



\printAffiliationsAndNotice{}  % leave blank if no need to mention equal contribution

\begin{abstract}
We consider the problem of adaptive Model Predictive Control (MPC) for uncertain linear-systems with additive disturbances and with state and input constraints. We present STT-MPC (Self-Tuning Tube-based Model Predictive Control), an online algorithm  that combines the certainty-equivalence principle and polytopic tubes. Specifically, at any given step, STT-MPC infers the system dynamics using the Least Squares Estimator (LSE), and applies a controller obtained by solving an MPC problem using these estimates. The use of polytopic tubes is so that, despite the uncertainties, state and input constraints are satisfied, and recursive-feasibility and asymptotic stability hold. In this work, we analyze the regret of the algorithm, when compared to an oracle algorithm initially aware of the system dynamics. We establish that the expected regret of STT-MPC does not exceed  $O(T^{1/2 + \epsilon})$, where $\epsilon \in (0,1)$ is a design parameter tuning the persistent excitation component of the algorithm. Our result relies on a recently proposed {\it exponential decay of sensitivity} property and, to the best of our knowledge, is the first of its kind in this setting. We illustrate the performance of our algorithm using a simple numerical example.
\end{abstract}

\section{Introduction}

The problem of optimal decision making for uncertain dynamical systems has been studied in both the control and the learning communities, and is referred to as stochastic optimal control, adaptive control, or reinforcement learning. A fundamental special case of this problem is the online (or adaptive) Linear Quadratic Regulator (LQR) which deals with the optimal control of systems with linear dynamics \cite{recht2019tour}. Early efforts in the control community yielded algorithms with asymptotic performance and stability guarantees \cite{aastrom1973self,lai1986}. Over the last decade, the problem has been revisited in the learning community (see e.g. \cite{abbasi2011regret,mania2019certainty, goel2022power, jedra2022minimal} and references therein), with the aim of devising algorithms with finite-time guarantees on the {\it regret}, defined as the difference between the cumulative cost of the learning algorithm and that of an oracle which has perfect knowledge of the system. 

All aforementioned studies deal with the simple LQR problem and cannot account for constraints on the system state and input. These constraints are important in applications as they capture both the inherent limitations (e.g position, velocity, and actuator limits) as well as limitations imposed due to safety or the cost of operating the system. Unfortunately, it is well known that directly solving a constrained LQR problem is in general intractable because of the infinite time horizon \cite{scokaert1998constrained}. A way to circumvent this issue is to use Model Predictive Control (MPC) which solves a finite-time constrained optimization problem in a receding horizon manner. 

Early work in the control community led to a better understanding of the connection between MPC and LQR and to algorithms with recursive feasibility and asymptotic stability guarantees \cite {mayne2000constrained}. In parallel, the tube MPC framework was developed to achieve the robust satisfaction of constraints in the presence of uncertainties \cite{mayne2005robust}. Recently, online MPC algorithms, combining tube MPC and adaptive strategies (to cope with the system uncertainty), have been proposed and shown to ensure recursive feasibility and asymptotic stability \cite{lorenzen2017adaptive, lu2019robust, lu2021robust}, and \cite{tranos2022self}. The learning community has also begun to investigate MPC, again with the aim of providing finite-time regret guarantees \cite{li2019online, yu2020power, zhang2021regret, lin2021perturbation, lin2022bounded}. A key limitation of these analyses is that they are valid only in the absence of constraints which goes against the main motivation of MPC \cite{bitmead1990adaptive}. Furthermore, most of these works assume to have access to (often perfect) system dynamics predictions. And those not making this assumption yield regret bounds having the cost of the oracle as an additive term. In turn, this term may scale linearly with time in the presence of constraints and additive disturbances.

In this work, we present STT-MPC, an adaptive MPC algorithm originally proposed in \cite{tranos2022self}. Inspired by the analysis pipeline of \cite{lin2022bounded}, we leverage the exponential decay of sensitivity property of the underlying finite-time optimization problem \cite{shin2022exponential} to provide upper bounds of the expected regret of the algorithm. Specifically, we show that this regret scales at most as  $T^{1/2 + \epsilon}$ where $\epsilon \in (0,1)$ is a design parameter tuning the persitent excitation component of the algorithm.

\medskip
\noindent
{\bf Notation.} For a time dependent vector $x_t$, we denote by $x_{k|t}$ its prediction at time $k+t$ given information at time $t$. For any two sets $\set{A}$ and $\set{B}$, we define their Minkowski sum as the set $\set{A}\oplus\set{B} := \{a+b: a\in \set{A}, \ b\in \set{B}\}$. We also define, for any constant $\lambda \geq 0$, the scaled set $\lambda \set{A} := \{\lambda a, \ a \in \set{A}\}.$ For any $d \in \NN$, $x \in \RR^d$, and $\epsilon > 0$, let $\set{B}(x,\epsilon)=\{y: \|y-x\|\le \epsilon \}$. Let $\set{B}=\set{B}(0,1)$. For any set $\set{S}$, and any $\varepsilon > 0$, there exists a polytope $\set{P}$ that is an outer approximation of $\set{S}$, i.e., $\set{S} \subset \set{P} \oplus \varepsilon \set{B}$. We refer to this polytope as the outer polyhedral approximation of $\set{S}$. Unless stated otherwise, we use $\norm{\cdot}$ to denote the Euclidean norm for vectors and the Frobenius norm for matrices.
\section{Problem Setting}

We consider the following discrete time, linear, time-invariant system:
\begin{align}
x_{t+1} =  A(\theta^\star) x_t + B(\theta^\star) u_t + w_t,
\label{eq:linear_system}
\end{align}
where $x_t, w_t\in \mathbb{R}^{d_x}$ and $u_t\in \mathbb{R}^{d_u}$. The state transition and state-action transition matrices $A(\theta^\star)$ and $B(\theta^\star)$ are initially unknown. The set of possible such matrices is parameterized by $\theta\in \mathbb{R}^{d_\theta}$ (here $\theta$ could well parameterize each entry of the matrices, in which case $d_\theta=d_x(d_x+d_u)$). To simplify the notation, for two possible parameters $\theta_1,\theta_2$, we define $\norm{\theta_1 - \theta_2}:= \max(\norm{A(\theta_1) - A(\theta_2)},\norm{B(\theta_1) - B(\theta_2)})$. We make the following assumptions.
\begin{assumption}[Parameter uncertainty]\label{ass1} The decision maker does not know $\theta^\star$, but knows that $\theta^\star\in\Theta_0$ where $\Theta_0$ is a convex polytope. Moreover, there exists  $\epsilon_0>0$ such that ${\cal B}(\theta^*,\epsilon_0)\subset \Theta_0$. 
\label{ass:parameter_set}
\end{assumption}
\begin{assumption}[Additive disturbance]
The sequence $(w_t)_{t\ge 0}$ is i.i.d, and for each $t\ge 0$, $w_t$ is zero-mean, isotropic, with support in the ball $\set{B}(0,3\sigma)$. Hence, $w_t$ is $\sigma^2$-sub-gaussian. Further define $\set{W}$, a convex polytope providing a conservative approximation of $\set{B}(0,3\sigma)$, i.e., $\set{B}(0,3\sigma) \subset \set{W}$.
\label{ass:noise}
\end{assumption}
\begin{assumption}[State and input constraints]
The set
\begin{align*}
\set{C} = \{(x,u) \in \RR^{d_x}\times \RR^{d_u}:Fx + Gu \leq \textbf{1}\}.
\end{align*}
is compact and contains the origin in its interior. Here, $F \in \RR^{d_c \times d_x}$ and $G \in \RR^{d_c \times d_u}$ define the state and input constraints respectively. The above inequality holds component-wise, and $\textbf{1}$ is the vector with all components equal to 1.
\label{ass:constraints}
\end{assumption}
\begin{assumption}[Stabilizing Controller]
There exists a known, robustly stabilizing feedback gain $K$ such that $A(\theta)+B(\theta)K$ is stable (i.e., $\rho(A(\theta)+B(\theta)K) <1$) for all $\theta \in \Theta_0$.
\label{ass:stabilizing_controller}
\end{assumption}

\section{Model Predictive Control}\label{sec:method}

We wish to minimize the long-term cost defined as $\lim\sup_{T\to\infty} \frac{1}{T}\sum_{t=0}^{T-1} \EE\left[x_t^\top Qx_t+u_t^\top Ru_t\right]$, through some positive semi-definite matrices $Q, R$. To this aim, we use MPC, with a receding horizon $N$. Specifically, at time $t$, given the current system state $x_t$ and the past observations used to derive an estimator $\theta_t$ of $\theta^\star$, we will identify a control policy $(u_{k|t})_{k=0,\ldots,N-1}$ minimizing the cost along a predicted system trajectory $(x_{k|t})_{k=0,\ldots,N}$. We use the well-known dual mode prediction paradigm \cite{kouvaritakis2016model} with the following predicted control sequence,
\begin{align*}%\label{eq:prediction_control}
u_{k|t} = \begin{cases}Kx_{k|t} + v_{k|t} \quad \forall k \in \{0,\dots,N-1 \}, \\ Kx_{k|t} \quad \forall k \geq N, \end{cases}
\end{align*} 
where $\{v_{0|t},\dots v_{N-1|t}\}$ are the optimization variables to be determined by the MPC. The resulting prediction dynamics will be for $k \in \{0,\dots,N-1\}$,
\begin{subequations}
\begin{align*}
x_{0|t} &= x_t, \\
x_{k+1|t} &= \Phi(\theta_t)x_{k|t} + B(\theta_t)v_{k|t},
\end{align*}
%\label{eq:prediction_dynamics}
\end{subequations}
where $\Phi(\theta):=A(\theta)+B(\theta)K$ for any $\theta\in \Theta_0$.

\subsection{LSE and persistent excitation}

Our algorithm starts with an initial parameter $\theta_0\in \Theta_0$, which is then updated using the LSE. We let $\theta_1=\theta_0$. For $t\ge 2$, the LSE enjoys the following explicit expression:
\begin{align*}
\hat{\theta}_t = {\left( \sum\limits_{k=0}^{t-2}x_{k+1} \begin{bmatrix}
x_k \\ u_k
\end{bmatrix}^\top
 \right) \left( \sum\limits_{k=0}^{t-2}\begin{bmatrix}
 x_k \\ u_k
 \end{bmatrix}
 \begin{bmatrix}
 x_k \\ u_k
 \end{bmatrix}^\top
 \right)^\dagger}.
 %\label{eq:LSE}
\end{align*}
For $t\ge t^\star(\delta)$ (to be defined later), we align our prediction parameter $\theta_t$ to $\hat{\theta}_t$. As shown in \cite{jedra2022minimal}, a finite-time analysis of the performance of the LSE is rather intricate but possible even if the feedback controller varies over time. The performance is tightly related to the minimal eigenvalue of the cumulative covariate matrix $\lambda_{\min}(\sum_{s=0}^{t-2}y_sy_s^\top)$ where $y_s=\begin{bmatrix}
x_s \\ u_s
\end{bmatrix}$. More precisely, for the LSE to lead to a good approximation of $\theta^\star$, we need to ensure that this eigenvalue grows with time. To this aim, we add an isotropic and bounded noise to the control input. This noise is represented by the random vector $\zeta_t$ taken to be the projection of $\xi_t$ on $\set{B}(0,3\sigma_t)$, where $\xi_t$ is i.i.d. according to a normal distribution, i.e., $\xi_t \sim \set{N}(0,\sigma_t^2 I_{d_u})$. The choice of $\sigma_t$ directly impacts the performance of the LSE but also the overall performance of the controller (a higher $\sigma_t$ means higher excitation and hence better LSE, but at the expense of a worse overall controller). Here we set $\sigma_t^2 = \sqrt{d_x}\sigma^2 t^{-\alpha}$ for some $\alpha\in (0,1)$. This ensures that (i) the LSE $\theta_t$ converges to $\theta^\star$ and (ii) the controller converges to that obtained through a classical tube-based MPC framework with known $\theta^\star$. In \cite{tranos2022self}, we have shown that the following {\it good} event ${\cal G}$ holds with probability at least $1-\delta$:
$$
{\cal G}=\left(\| \hat{\theta}_t-\theta^\star\|\le \epsilon_t, \ \forall t\ge t^\star(\delta)\right),
$$
where $t^\star(\delta)=c_1+c_2\log(1/\delta)$ and $\epsilon_t^2 = c_3 \log(t/\delta)/t^{1-\alpha}$ for some positive constants $c_1, c_2, c_3$. Next, we define $\Delta _t$ as an outer polyhedral approximation of $\set{B}(\theta_t,2\epsilon_t)$. We further recursively define the {\it uncertainty sets} as follows: $\Theta_t=\Theta_{t-1}\cap \Delta_t$ for all $t\ge t^\star(\delta)$ and $\Theta_t=\Theta_0$ for $t<t^\star(\delta)$. By construction, the true parameter $\theta^\star$ belongs to the interior of $\Theta_t$ with high probability in the following sense:

\begin{lemma}
Under event ${\cal G}$, $\set{B}(\theta^\star,\epsilon_t) \subset \Theta_t$ for all $t\ge 1$.
\label{prop:inclusion}
\end{lemma}
%\proof See \cite{tranos2022self}
\proof For $t< t^\star(\delta)$ we have $\Theta_t = \Theta_0$ and the result holds by Assumption \ref{ass1}. Let $t\ge t^\star(\delta)$. We show that $\set{B}(\theta^*,\epsilon_t) \subset \Delta_t$ and $\set{B}(\theta^*,\epsilon_t) \subset \Theta_{t-1}$. For the first, we have for all $\theta \in \set{B}(\theta^*,\epsilon_t)$, using the triangle inequality: $\norm{\theta-\theta_t} \leq \norm{\theta-\theta^*} + \norm{\theta_t-\theta^*} \leq 2\epsilon_t$, where the second inequality holds under ${\cal G}$. Thus $\theta \in \set{B}(\theta_t,2\epsilon_t)$ and so $\set{B}(\theta^*,\epsilon_t) \subset \set{B}(\theta_t,2\epsilon_t) \subset \Delta_t$. We prove $\set{B}(\theta^*,\epsilon_t) \subset \Theta_{t-1}$ by induction. Assume that $\set{B}(\theta^*,\epsilon_t) \subset \Theta_{t-1}$. Then we show that $\set{B}(\theta^*,\epsilon_{t+1}) \subset \Theta_{t}$. Let $\theta$ such that $\|\theta-\theta^\star\|\le \epsilon_{t+1}$. Then $\|\theta-\theta_t\|\le \norm{\theta-\theta^*} + \norm{\theta_t-\theta^*}\le \epsilon_{t+1}+\epsilon_t$. This implies that $\theta\in \Delta_t$. In addition, $\theta\in \Theta_{t-1}$. Indeed, $\|\theta-\theta^\star\|\le \epsilon_{t+1}\le \epsilon_t$, and we conclude using the induction assumption. Hence $\theta\in \Theta_t$.\ep

\begin{algorithm}[tb]
\caption{STT-MPC}
\label{alg:main}
\begin{algorithmic}
\STATE {\bfseries Input:} Initial state $x_0$; confidence $\delta$; estimate $\theta_0$; uncertain parameter set $\Theta_0$
\STATE Find stabilizing matrix $K$ for all $\theta\in \Theta_0$
\STATE Compute matrices $T$ and $H_c$
\FOR{$t=1,\dots,\set{T}$}
\IF{$t < t^\star(\delta)$}
\STATE Set $\theta_t \gets \theta_{t-1}$ and  $\Theta_t \gets \Theta_{t-1}$
\ELSE
\STATE Update $\theta_t\gets\hat{\theta}_t$
\STATE Compute $\Delta_t$ and set $\Theta_t \gets \Theta_{t-1} \cap \Delta_t$
\ENDIF
\STATE Compute matrices $H_t^{(j)}$
\STATE Solve $\set{P}_N(x_t,\theta_{\rho(t)})$ and obtain $v^\pi_{0|t}$
\STATE Apply $u_t = Kx_t + v^\pi_{0|t} + \zeta_t$
\ENDFOR 
\end{algorithmic}
\end{algorithm}

\subsection{Polytopic tubes and associated linear constraints}

With the considered control inputs, the system can be rewriten as
\begin{equation*}%\label{eq:newdyn}
x_{t+1}=\Phi(\theta^\star)x_t+B(\theta^\star)(v_{0|t}+\zeta_t)+w_t.
\end{equation*}
We apply a tube-based approach, and at time $t$, we build a polytopic tube based on:
\begin{itemize}
    \item[(i)] $\Theta_t$, encoding the uncertainty about $\theta^\star$. We denote by $m$ the number of vertices of $\Theta_t$, and the vertices themselves by $\theta_t^{(j)}$, $j=1,\ldots,m$.
    \item[(ii)] A polytope $\bar{\set{W}}_t$, handling the uncertainty due to the {\it noise} $B(\theta^\star)\zeta_t+w_t$, including that due to the persistent excitation. To define $\bar{\set{W}}_t$, let $\bar{B}_t = \max_{\theta \in \Theta_t}\norm{B(\theta)}_2$. We define $\set{Z}_t$ as the outer polyhedral approximation to $\set{B}(0,3\sigma_t)$, and $\bar{\set{W}}_t = \set{W} \oplus \bar{B}_t\set{Z}_t$.
\end{itemize}
\noindent
We define the state tube cross sections as the sets: for $k=0,\ldots, N$,
\begin{align*}
X_{k|t} = \{x: Tx \leq \alpha_{k|t} \},
\end{align*}
where the matrix  $T \in \RR^{d_\alpha \times d_x}$ is chosen such that, for some $\lambda \in [0,1)$, the set $\{x: Tx \leq \mathbf{1} \}$ is $\lambda$-contractive with respect to the system $x_{t+1} = \Phi(\theta)x_t$ for all $\theta\in \Theta_0$. This property is needed to ensure the robust positive invariance of $X_{N|t}$ (see Lemma 5.7 in \cite{kouvaritakis2016model}). To derive the associated linear constraints, we apply a standard result to ensure inclusion of polyhedra (see Proposition 3.31 in \cite{blanchini2008set}). More precisely, for any $j=1,\ldots,m$, we have $X_{k|t} \subseteq \{x: \Phi(\theta_t^{(j)}) x + B(\theta_t^{(j)})v_{k|t} + w \in X_{k+1|t}\}$ for all $w\in \bar{\set{W}}_t$ if there exists $H_t^{(j)} \geq 0$ such that:
\begin{subequations}
\begin{align*}
&H_t^{(j)}T = T\Phi(\theta_t^{(j)}), \\
&H_t^{(j)}\alpha_{k|t} + TB(\theta_t^{(j)})v_{k|t} + \bar{w}_t\leq \alpha_{k+1|t},
\end{align*}
\end{subequations}
where $\bar{w}_t$ is such that $[\bar{w}_t]_i= \max_{w \in \bar{\set{W}}_t}[Tw]_i$ for $i\in \{1,\dots, d_{\alpha}\}$.
\noindent
Similarly, we have $X_{k|t} \subseteq \{x: (F+GK) x + G(v_{k|t} + \zeta)\leq \textbf{1}\}$ for all $\zeta \in \set{Z}_t$, if there exists $H_c \geq 0$, such that:
\begin{subequations}
\begin{align*}
&H_c T = F+GK \\
&H_c \alpha_{k|t} + G v_{k|t} + \bar{\zeta}_t \leq \textbf{1},
\end{align*}
\end{subequations}
where $\bar{\zeta}_t$ is such that $[\bar{\zeta}_t]_i = \max_{\zeta \in \set{Z}_t} [G\zeta]_i$ for $i\in \{1,\dots, d_c\}$.
\begin{remark}
For simplicity, we have chosen $\bar{w}_t$ and $\bar{\zeta}_t$ so as to be conservative with respect to the noise we apply. However, notice that we could well pre-sample the noise sequence $(\zeta_{t})_{t\ge 0}$ and use it as part of our predictions.
\end{remark}
The matrices $H_t^{(j)}$, and $H_c$ are chosen such that for all $i \in \{ 1,\dots, d_\alpha \}$ and for all $j \in \{1,\dots,m\}$,
\begin{subequations}
\begin{align*}
\big( H_t^{(j)} \big)_i &= \arg \min_h \{\textbf{1}^\top h: h^\top T = (T)_i \Phi(\theta_t^{(j)}),\ h\geq 0 \},
%\label{eq:H^j}
\end{align*}
and for all $i \in \{1,\dots, d_c\}$,
\begin{align*}
\big( H_c \big)_i &= \arg \min_h \{\textbf{1}^\top h: h^\top T = (F+GK)_i,\ h\geq 0 \}.
%\label{eq:H_c}
\end{align*}
\end{subequations}
Finally, we have the terminal conditions
\begin{align*}
H_t^{(j)} \alpha_N + \bar{w}_t &\leq \alpha_N, \\
H_c\alpha_N + \bar{\zeta}_t &\leq \mathbf{1}.
\end{align*}

\subsection{The tube MPC problem}

Let $\boldsymbol{v}_t = \{v_{0|t},\dots,v_{N-1|t}\}$ and $\boldsymbol{\alpha}_t = \{\alpha_{0|t},\dots, \alpha_{N|t}\}$. The resulting tube MPC problem, denoted as $\set{P}_N(x_t,\theta)$ for any parameter $\theta$ is:
\begin{align}
&\underset{\boldsymbol{v}_t,\boldsymbol{\alpha}_t}{\textnormal{minimize}} \sum_{k=0}^{N-1}\left(\ x_{k|t}^\top Q x_{k|t} +  v_{k|t}^\top R v_{k|t}\right) + x_{N|t}^\top P(\theta) x_{N|t} \nonumber \\
&\textnormal{subject to}, \textnormal{for all } j=1,\dots,m, \textnormal{and } k = 0,\dots,N-1: \nonumber \\
&\textnormal{initial constraints:} \nonumber \\
&\quad x_{0|t} = x_t, \label{eq:initial1} \\
&\quad Tx_{0|t} \leq \alpha_{0|t}, \label{eq:initial2} \\
&\textnormal{system constraints}, \nonumber \\
&\quad x_{k+1|t} = \Phi(\theta)x_{k|t} + B(\theta)v_{k|t}, \label{eq:system2} \\
&\textnormal{tube constraints}, \nonumber \\
&\quad H_t^{(j)}\alpha_{k|t} + TB(\theta^{(j)})v_{k|t} + \bar{w}_t \leq \alpha_{k+1|t}, \label{eq:tube1}\\
&\quad H_c\alpha_{k|t} + G v_{k|t} + \bar{\zeta
}_t \leq \mathbf{1}, \label{eq:tube2}\\
&\textnormal{terminal constraints:} \nonumber \\
&\quad H_t^{(j)}\alpha_{N|t} + \bar{w}_t \leq \alpha_{N|t}, \label{eq:terminal1} \\
&\quad H_c\alpha_{N|t} + \bar{\zeta
}_t\leq \mathbf{1} \label{eq:terminal2}.
\end{align}
where $P(\theta)$ is obtained by solving the Lyapunov equation:
\begin{align}
P(\theta)-\Phi(\theta)^\top P(\theta)\Phi(\theta) = Q+K^\top R K
\label{eq:lyapunov}
\end{align}

\begin{remark}
    The oracle tube-MPC problem $\set{P}_N(x_t,\theta^\star)$ is similarly defined by setting $\theta = \theta^\star$, $\bar{w}_t = \bar{w}_t^\star$ (with $[\bar{w}_t^\star]_i= \max_{w \in \set{W}}[Tw]_i$, for $i\in \{1,\dots, d_{\alpha}\}$), and $m = 1$. The later can be achieved by Algorithm \ref{alg:main} by removing redundant constraints at every iteration by solving a linear program or one of the methods outlined in \cite{paulraj2010comparative}.
\end{remark}

The feasibility of the above problem depends on whether $\theta^\star\in \Theta_t$, i.e., on the event ${\cal G}$. However, for the sake of the analysis, we would like to ensure that the problem the algorithm solves in each step is always feasible (with probability 1). This is necessary for establishing that ${\cal G}$ holds with probability at least $1-\delta$ (Theorem 1 of \cite{tranos2022self}) as well as for our regret analysis. To ensure that the tube MPC problem is always feasible, i.e., even when $\set{G}$ does not occur, we instead solve $\set{P}_N(x_t,\theta_{\rho(t)})$\footnote{More precisely, the problem is obtained by replacing $\theta_t$ by $\theta_{\rho(t)}$, $H_t^{(j)}$ by $H_{\rho(t)}^{(j)}$ for all $j$ in $\set{P}_N(x_t,\theta_t)$.} in Algorithm \ref{alg:main} where,
\begin{align*}
    \rho(t) := \max\{\tau \leq t: \set{P}_N(x_t,\theta_\tau) \ \textnormal{is feasible} \}.
\end{align*}
Essentially, in the unlikely event that $\set{P}_N(x_t,\theta_t)$ is not feasible, we instead solve $\set{P}_N(x_t,\theta_\tau)$ using the latest estimates $\theta_\tau$ and $\Theta_\tau$ for which the problem is feasible. This modification is of little practical consequence as we can (and typically do) choose $\delta$ to be very small.

With this, we may state the recursive feasibility property of STT-MPC (Theorem 2 in \cite{tranos2022self}):
\begin{theorem}
If the optimization problem $\set{P}_N(x_0,\theta_0)$ is feasible for initial state $x_0 \in \RR^{d_x}$ and parameter $\theta_0 \in \Theta_0$, then, for all $t>0$, \\
(i) under event $\set{G}$, the problem $\set{P}_N(x_t,\theta_t)$ is feasible;\\
(ii) the problem $\set{P}_N(x_t,\theta_{\rho(t)})$ is feasible.
\label{thm:recursive_feasibility} 
\end{theorem}

\section{Regret Analysis}\label{sec:regret_analysis}

In this section, we present our main result, an upper bound on the expected regret of STT-MPC. To state our main theorem, we introduce the following notation. Let $\Pi$ be the set of all tube-MPC algorithms (which are parameterized by $\theta$). For an algorithm $\pi \in \Pi$, define $(x_0^\pi, u_0^\pi, \dots ,x_t^\pi, u_t^\pi)$ as the sequence of states and control inputs generated under $\pi$, with $x_0^\pi = x_0$. We also define by $v_t^\pi(x,\theta) := v_{0|t}^\pi(x,\theta)$, the first element of the solution of the optimization problem $\set{P}_N(x,\theta)$. We denote by $z_t^\pi(x)$ the state of the system \eqref{eq:linear_system} at time $t$ under algorithm $\pi$ given initial state $x$, and we denote by $y_t^\pi(x)$, the associated output of $\pi$. Let $\pi_\star \in \Pi$ be the oracle tube-MPC problem $\set{P}_N(x,\theta^\star)$. We simplify the notation by replacing $\pi_\star$ by $\star$, e.g., $x_t^\star$, $v_t^\star$ and so on. We define the regret $R_T(\pi)$ of an algorithm $\pi$ as:
\begin{align*}
R_T(\pi) = \sum_{t=0}^{T-1} (\norm{x_t^\pi}^2_Q + \norm{u_t^\pi}^2_R) - \sum_{t=0}^{T-1}(\norm{x_t^{\star}}^2_Q + \norm{u_t^{\star}}^2_R).
\end{align*}

\begin{theorem}
\label{thm:regret}
    The regret of $\pi$ = STT-MPC with confidence $\delta$ and persistent excitation parameter $\alpha := 2\epsilon$, with $\epsilon\in (0,1)$ satisfies for all $T \geq 1$, with probability at least $1-\delta$,
$$
R_T(\pi) \le C\left( (\log(T/\delta)^{1/2}+1)T^{1/2+\epsilon}+\log(1/\delta)+1\right),
$$
for some universal constant $C>0$. Now when the confidence is set to $1/T^2$, we get   
    \begin{align*}
        \EE[R_T(\pi) \le C' \log(T)^{1/2}T^{1/2+\epsilon},
    \end{align*}
    where $C'$ is a universal constant.
\end{theorem}

\proof We first express the regret as the sum of differences between the states and inputs,
\begin{align*}
R_T(\pi) =&\sum_{t=0}^{T-1}\left( {x_t^\pi}^\top Q (x_t^\pi - x_t^\star) + (x_t^\pi - x_t^\star)^\top Q x_t^\star\right) \\
&+\sum_{t=0}^{T-1}\left( {u_t^\pi}^\top R (u_t^\pi - u_t^\star) + (u_t^\pi - u_t^\star)^\top R u_t^\star \right)
\\
\leq& \sum_{t=0}^{T-1} (\norm{x_t^\pi} + \norm{x_t^\star})\norm{Q}\norm{x_t^\pi - x_t^\star} \\
&+\sum_{t=0}^{T-1}(\norm{u_t^\pi} + \norm{u_t^\star})\norm{R}\norm{u_t^\pi - u_t^\star} \\
=& \bar{Q}\sum_{t=0}^{T-1}\norm{x_t^\pi - x_t^\star} + \bar{R}\sum_{t=0}^{T-1}\norm{u_t^\pi - u_t^\star},
\end{align*}
%where the first inequality follows from Cauchy-Schwartz inequality, and the second by defining $\bar{Q} := 2\norm{Q}\max\{\norm{x_t}\}$ and $\bar{R} := 2\norm{R}\max\{\norm{u_t}\}.$ Here, $\max\{\norm{x_t}\}$ and $\max\{\norm{u_t}\}$ can be inferred from the bounds of constraint set $\set{C}$ which is compact by Assumption \ref{ass:constraints}. Note that $(x_t,u_t) \in \set{C}$ is guaranteed by the recursive feasibility of STT-MPC (Theorem \ref{thm:recursive_feasibility}).
where the first inequality follows from Cauchy-Schwartz inequality, and the second by defining $\bar{Q} := 2\norm{Q}\bar{x}$ and $\bar{R} := 2\norm{R}\bar{u}$ with $\bar{x}=\max_{(x,u)\in {\cal C}}\{\norm{x}\}$ and $\bar{u}=\max_{(x,u)\in {\cal C}}\{\norm{u}\}$. Note that $\bar{Q}$ and $\bar{R}$ are well defined since the constraint set $\set{C}$ is compact by Assumption \ref{ass:constraints}. Further note that $(x_t,u_t) \in \set{C}$ is guaranteed by the recursive feasibility of STT-MPC (Theorem \ref{thm:recursive_feasibility}).

We seek to upper bound the quantities $\norm{x_t^\pi - x_t^\star}$ and $\norm{u_t^\pi - u_t^\star}$ in terms of $\norm{\theta_t - \theta^\star}$. To this end, we make use of the following theorem, presented originally in \cite{shin2022exponential} (Theorem 4.5) for graph-structured nonlinear optimization problems, and specialized to constrained nonlinear systems by \cite{lin2022bounded} (Theorem H.1.).

\begin{theorem}

For any tube-MPC algorithm $\pi \in \Pi$, there exists $R > 0$ such that, for $x_i, x_i' \in B(x^\star_i, R)$, where $i \in \{0, \dots, t\}$, the following perturbation bounds hold:
\begin{align}
    \norm{z_t^\pi(x_i, \theta) - z_t^\pi(x_i', \theta)} \leq q_1(t-i) \norm{x_i - x_i'},
    \label{eq:state_bound}
\end{align}
and
\begin{align}
    \norm{v_t^\pi(x_i, \theta) - v_t^\pi(x_i, \theta')} \leq q_2(t-i) \norm{\theta - \theta'},
    \label{eq:input_bound}
\end{align}
where, for $j \in \{1,2\}$, $q_j(t)$ is a function such that $\lim_{t \to \infty}\lim q_j(t) = 0$ and $\sum_{t=0}^\infty q_j(t) \leq C_j$, for some constant $C_j \geq 1$.

\end{theorem}

The bound \eqref{eq:input_bound} can be applied directly to the input error (along the trajectory induced by $\pi$) $e_t := \norm{v_t^\pi - v_t^\star(x_t^\pi)}$, provided that $x_t^\pi \in \set{B}(x_t^\star, R)$. Therefore, we show inductively that if $e_t \leq R /(\norm{B(\theta^\star)}C_1^2)$ then $x_t^\pi \in \set{B}(x_t^\star, R / C_1)$. First, observe that
\begin{align}
    \norm{x^\pi_{t} - z^\star_{t}(x^\pi_{t-1})} &= \norm{B(\theta^\star)(v_{t-1}^\pi - v_{t-1}^\star(x_{t-1}^\pi))} \nonumber \\ 
    & \leq \norm{B(\theta^\star)}e_{t-1}, \label{eq:one-step-bound}
\end{align}
and so the condition holds for the base case of $t = 0$, since $x_0^\pi = x_0^\star$. Let our induction hypothesis be that it holds for $0, \dots, t-1$. We will have
\begin{align*}
&\norm{x_t^\pi - x_t^\star} \\
&= \norm{x_t^\pi - z_t^\star(x_0)} \\
& \leq \norm{x_t^\pi - z_t^\star(x_{t-1})} + \sum_{i=1}^{t-1}\norm{z_t^\star(x_{t-i}) - z_t^\star(x_{t-i-1})} \\
& = \norm{x_t^\pi - z_t^\star(x_{t-1})} + \sum_{i=1}^{t-1}\norm{z_t^\star(x_{t-i}) - z_t^\star(z_{t-1}^\star(x_{t-i-1}))} \\
& \leq \norm{x_t^\pi - z_t^\star(x_{t-1})} + \sum_{i=1}^{t-1}q(i)\norm{x_{t-i} - z_{t-i}^\star(x_{t-i-1})} \\
& \leq \sum_{i=0}^{t-1}q_1(i)\norm{x_{t-i} - z_{t-i}^\star(x_{t-i-1})} \\
& \leq \norm{B(\theta^\star)}\sum_{i=0}^{t-1}q_1(t)e_{t-i-1},
\end{align*}
where the first inequality is a straightforward application of the triangle inequality, the second equality follows by the definition of $z^\star_t$, and the third inequality is a direct application of \eqref{eq:state_bound} (noting that $x_{t-i}^\pi \in \set{B}(x_{t-i}^\star, R / C_1)$ by the induction hypothesis). The last inequality follows from \eqref{eq:one-step-bound}.

Now if we substitute $e_{t-i} \leq  R / (\norm{B(\theta^\star)} C_1^2)$, we obtain
\begin{align*}
    \norm{x_t^\pi - x_t^\star} \leq \norm{B(\theta^\star)} \frac{R}{\norm{B(\theta^\star)} C_1^2}\sum_{i=0}^{t-1}q_1(t) \leq \frac{R}{C_1},
\end{align*}
which proves the induction step.

Proceeding similarly, we obtain the following bound,
\begin{align*}
&\norm{v_t^\pi - v_t^\star} \\
&= \norm{v_t^\pi - y_t^\star(x_0)} \\
& \leq \norm{v_t^\pi - y_t^\star(x_{t-1})} + \sum_{i=0}^{t-1}\norm{v_t^\star(x_{t-i}) - y_t^\star(x_{t-i-1})} \\
& \leq \norm{v_t^\pi - v_t^\star(x_{t-1})} + \sum_{i=0}^{t-1}q_1(i)\norm{x_{t-i} - z_{t-i}^\star(x_{t-i-1})} \\
& \leq e_t + \norm{B(\theta^\star)}\sum_{i=0}^{t-1}q_1(t)e_{t-i-1},
\end{align*}
where again, we have used the triangle inequality for the first inequality, and the bound \eqref{eq:state_bound} for the second.
Finally, we have
\begin{align*}
    \norm{u^\pi_t - u^\star_t} &= \norm{v^\pi_t + \zeta_t - v^\star_t} \\
    &\leq \norm{v^\pi_t - v^\star_t} + \norm{\zeta_t}.
\end{align*}
With these bounds in hand, we first account for the regret up to time $t^\star(\delta)$ while the event $\set{G}$ holds. Note that a crude bound is sufficient to show that it is finite:
\begin{align*}
R_{t^\star(\delta)}(\pi)_\set{G} & \leq \bar{Q}\sum_{t=0}^{t^\star(\delta)}\norm{x_t^\pi - x_t^\star} + \bar{R}\sum_{t=0}^{t^\star(\delta)}\norm{u_t^\pi - u_t^\star} \\
& \leq (2\bar{Q}\bar{x} +  2\bar{R}\bar{u})t^\star(\delta) \\
& \leq \bar{C}_2 + \bar{C}_3 \log(1 / \delta).
\end{align*}
Where $\bar{C}_2 = (2\bar{Q}\bar{x} +  2\bar{R}\bar{u})c_2$ and $\bar{C}_3 = (2\bar{Q}\bar{x} +  2\bar{R}\bar{u})c_3$.

We then bound the regret from $t^\star(\delta)$ onward under the event $\set{G}$:
\begin{align*}
&(R_T(\pi) - R_{t^\star(\delta)}(\pi))_\set{G} \\
&\leq \bar{Q}\sum_{t=t^\star(\delta)}^{T-1}\norm{x_t^\pi - x_t^\star} + \bar{R}\sum_{t=t^\star(\delta)}^{T-1}\norm{u_t^\pi - u_t^\star} \\
&\leq\bar{Q}\sum_{t=0}^{T-1}\norm{x_t^\pi - x_t^\star} + \bar{R}\sum_{t=0}^{T-1}\norm{v^\pi_t - v^\star_t} + \norm{\zeta_t}\\
&\leq \bar{Q}\sum_{t=0}^{T-1}\norm{B(\theta^\star)}e_{t-1} \\
&+ \bar{R}\sum_{t=0}^{T-1}(e_t + \norm{B(\theta^\star)}\sum_{i=0}^{t-1}q_1(t)e_{t-i-1}) + 3\sigma_t\\
&\leq (\bar{Q} + \bar{R}C_1)\norm{B(\theta^\star)}\sum_{t=0}^{T-1}e_t + 3\sigma_t\\
&\leq (\bar{Q} + \bar{R}C_1)\norm{B(\theta^\star)}\sum_{t=0}^{T-1}q_2(0)\norm{\theta_t - \theta^\star} + 3\sigma_t \\
&\leq \bar{C} \sum_{t=0}^{T-1} \epsilon_t + 3\sigma_t \\
&\leq \bar{C}\sqrt{c_3} \sum_{t=0}^{T-1} \log(t/\delta)^{1/2}t^{-(1-\alpha)/2} + 3d_x^{1/4}\sigma \sum_{t=0}^{T-1} t^{-\alpha/2} \\
&\leq \frac{2\bar{C}\sqrt{c_3}}{\alpha + 1}\log(T/\delta)^{1/2}T^{(\alpha+1)/2} 
+\frac{6d_x^{1/4}}{\alpha+1}\sigma T^{(\alpha+1)/2},
\end{align*}
with $\bar{C} = (\bar{Q} + \bar{R}C_1)\norm{B(\theta^\star)}C_1$.

Putting it together, we have under the event $\set{G}$:
\begin{align*}
R_T(\pi)_\set{G} =& (R_T(\pi) - R_{t^\star(\delta)}(\pi))_\set{G} + R_{t^\star(\delta)}(\pi)_\set{G} \\
\leq& \frac{2\bar{C}\sqrt{c_3}}{\alpha + 1}\log(T/\delta)^{1/2}T^{(\alpha+1)/2} \\
&+ \frac{2d_x^{1/4}}{\alpha+1}\sigma T^{(\alpha+1)/2} 
+ \bar{C}_2\log(1 / \delta) + \bar{C}_1.
\end{align*}
To obtain a bound on the expected regret, first note that
\begin{align*}
  \EE[R_T(\pi)] = (1-\delta) R_T(\pi)_{\set{G}} + \delta R_T(\pi)_{\neg\set{G}}
\end{align*}
where $R_T(\pi)_{\neg\set{G}}$ is the regret when the event $\set{G}$ does not hold and can be taken to be linear in time with some constant $c$. This follows again from the recursive feasibility of STT-MPC (Theorem \ref{thm:recursive_feasibility}), which ensures that $x_t^\pi$ and $u_t^\pi$ are bounded in the set $\set{C}$.  Letting $\delta = 1/T^2$ leads to
\begin{align*}
  \EE[R_T(\pi)] &= (1-\delta) R_T(\pi)_{\set{G}} + \delta R_T(\pi)_{\neg\set{G}} \\
  &= R_T(\pi)_{\set{G}} +\frac{1}{T^2}R_T(\pi)_{\set{G}} + \frac{c}{T} \\
  &\lesssim \frac{2\sqrt{3}\bar{C}\sqrt{c_3}}{\alpha + 1}\log(T)^{1/2}T^{(\alpha+1)/2}.
\end{align*}
Here, the notation $\lesssim$ hides the universal constant such that the inequality holds. \ep

\section{Numerical Example} \label{sec:numerical}
We illustrate the performance of STT-MPC using the following second-order linear system:
\begin{equation}
    A = \begin{bmatrix}
    0.6 & 0.2\\-0.1 & 0.4
    \end{bmatrix},\quad
    B = \begin{bmatrix}
    1\\0.6
    \end{bmatrix},
    \quad \sigma=0.01,
\end{equation}
The initial state is $x_0=(6,3)$, and we consider $\Theta_0$ to be a $6$-dimensional hypercube centered on $\theta_0=\begin{bmatrix}0.57&0.17 &-0.12 & 0.42 & 0.95 & 0.65\end{bmatrix} $ with side length $0.14$. Consequently, the resulting stabilizing feedback gain is $K=\begin{bmatrix} -0.426 & -0.290\end{bmatrix}$. We consider the following state and input constraints: $[x_t]_1 \geq -0.15, [x_t]_2 \geq -1.1$ and $u_t \leq 0.5$. We inject a persistent excitation signal with a standard deviation of $\sigma_t = \sqrt{2}\sigma (t+1)^{-\alpha}$.

The matrix $T$ is computed according to the relation (5.98) in \cite{kouvaritakis2016model}, with $\lambda=0.999$. We consider $Q=I_{d_x}$, $R=I_{d_u}$, and, for simplicity, we let the worst case noise realization be $\bar w_t = \bar w_0$, for all $t\geq 1$.
All the simulations were performed in Python 3.9, using the CVXPY library \cite{diamond2016cvxpy} and the MOSEK solver.

\begin{figure}[t]
    \centering
    \includegraphics[width=\linewidth]{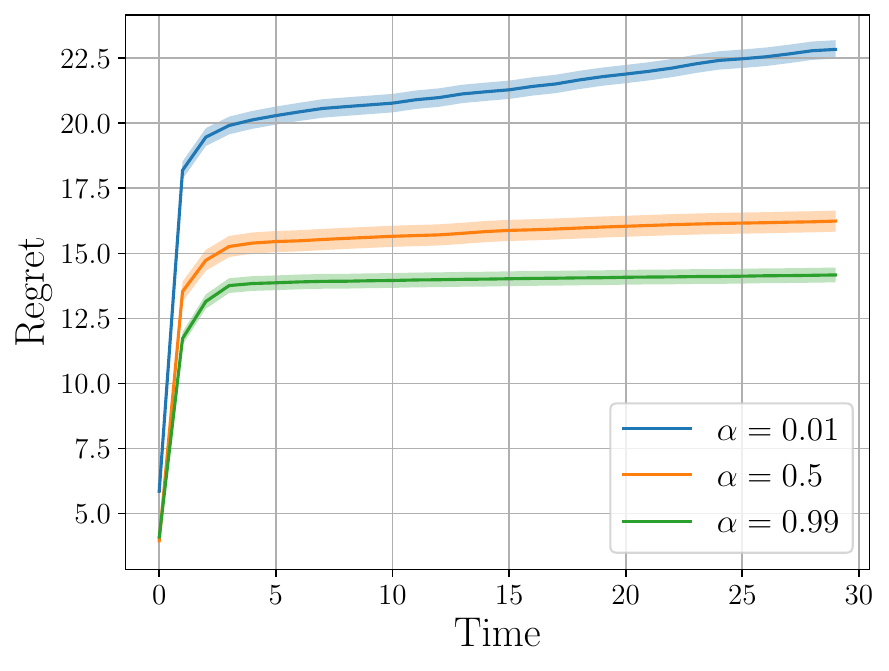}
    \caption{Regret vs. time of \textsc{STT-MPC} averaged over 100 runs (the shaded area corresponds to the standard error of the mean). Each curve corresponds to a different decay rate of the persistent excitation signal $\zeta_t$.}
    \label{fig:regret}
\end{figure}

In Figure \ref{fig:regret}, we present the regret of \textsc{STT-MPC} versus the oracle tube-MPC algorithm the cases where $\alpha$ is $0.01, 0.5,$ and $0.99$. For all three cases, the scaling is logarithmic and is thus over-estimated by the bound of Theorem \ref{thm:regret}. Note the lower the value of $\alpha$, the slower the decay rate of the excitation, which improves the transient estimation accuracy but negatively impacts the performance of the controller. The result suggests that the LSE converges very quickly to the true parameter (already before $t = 5$) even for high values of $\alpha$, and so the performance of the controller (rather, the decay rate of the excitation signal) dominates the regret.

Overall the logarithmic scaling is surprising as it is better than the $\sqrt{T}$ bound shown both theoretically and experimentally by \cite{jedra2022minimal} in the case of the \textsc{LQR}. We conjecture that the presence of constraints has a benign property on the regret and that it should be possible to tighten our upper bound to also scale as $\log(T)$.

\section{Conclusions} \label{sec:conclusion}

We proposed STT-MPC which combines least-squares estimaton with a polytopic tube-based MPC method to ensure robust constraint satisfaction while learning the system dynamics. Persistent excitation is ensured by injecting a truncated noise signal which decays at a rate $t^{-\alpha}$, with $\alpha > 0$ controlling the trade-off between (transient) estimation accuracy and controller performance. Importantly, we asymptotically recover the performance of the oracle tube-based MPC which has full knowledge of the dynamics.

We provided guarantees on the expected regret of our proposed algorithm by leveraging performance bounds on the LSE, the exponential decay of sensitivity property of the optimization problem, and the recursive feasibility of our algorithm. We show theoretically that the expected regret of STT-MPC scales at a rate of $T^{1/2 + \epsilon}$ with $\epsilon \in (0,1)$ and also demonstrated its performance via a numerical example.

The logarithmic rate demonstrated in simulation suggests that our bound can be further tightened, and we will investigate this further in future work.

\bibliography{RL.bib}
\bibliographystyle{icml2023}

\end{document}